\documentclass[aps,superscriptaddress,preprint]{revtex4-1}
\usepackage[latin9]{inputenc}
\usepackage{bm}
\usepackage{amsmath}
\usepackage{amssymb}
\usepackage{graphicx}

\begin{document}

\title{Concomitant enhancement of longitudinal spin Seebeck effect with
thermal conductivity}

\author{Ryo Iguchi}
\email{iguchi@imr.tohoku.ac.jp}

\affiliation{Institute for Materials Research, Tohoku University, Sendai 980-8577,
Japan}

\author{Ken-ichi Uchida}

\affiliation{Institute for Materials Research, Tohoku University, Sendai 980-8577,
Japan}

\affiliation{National Institute for Materials Science, Tsukuba 305-0047, Japan}

\affiliation{PRESTO, Japan Science and Technology Agency, Saitama 332-0012, Japan}

\affiliation{Center for Spintronics Research Network, Tohoku University, Sendai
980-8577, Japan}

\author{Shunsuke Daimon}

\affiliation{Institute for Materials Research, Tohoku University, Sendai 980-8577,
Japan}

\affiliation{WPI Advanced Institute for Materials Research, Tohoku University,
Sendai 980-8577, Japan}

\author{Eiji Saitoh}

\affiliation{Institute for Materials Research, Tohoku University, Sendai 980-8577,
Japan}

\affiliation{Center for Spintronics Research Network, Tohoku University, Sendai
980-8577, Japan}

\affiliation{WPI Advanced Institute for Materials Research, Tohoku University,
Sendai 980-8577, Japan}

\affiliation{Advanced Science Research Center, Japan Atomic Energy Agency, Tokai
319-1195, Japan}
\begin{abstract}
We report a simultaneous measurement of a longitudinal spin Seebeck
effect (LSSE) and thermal conductivity in a Pt/${\rm Y_{3}Fe_{5}O_{12}}$
(YIG)/Pt system in a temperature range from 10 to 300 K. By directly
monitoring the temperature difference in the system, we excluded thermal
artifacts in the LSSE measurements. It is found that both the LSSE
signal and the thermal conductivity of YIG exhibit sharp peaks at
the same temperature, differently from previous reports. The maximum
LSSE coefficient is found to be $S{\rm _{LSSE}}>10\ \mu{\rm V/K}$,
one-order-of magnitude greater than the previously reported values.
The concomitant enhancement of the LSSE and thermal conductivity of
YIG suggests the strong correlation between magnon and phonon transport
in the LSSE. 
\end{abstract}
\maketitle
A spin counterpart of the Seebeck effect, the spin Seebeck effect
(SSE) \citep{Uchida:2008cc}, has attracted much attention from the
viewpoints of fundamental spintronic physics \citep{Bauer:2012fq,MaekawaSadamichi:2013ik,maekawa2012spin}
and future thermoelectric applications \citep{Uchida:2016jo,Kirihara:2012jq,boona2014spin}.
The SSE converts temperature difference into a spin current in a magnetic
material, which can generate electrical power by attaching a conductor
with spin\textendash orbit interaction \citep{Uchida:2008cc,Uchida:2016jo}.
The SSE originates from thermally-excited magnons, and it appears
even in magnetic insulators. In fact, after the pioneer work by Xiao
\textit{et al.} \citep{Xiao:2010iy}, the SSE has been discussed in
terms of the thermal non-equilibrium between magnons in a magnetic
material and electrons in an attached conductor. Recent experimental
and theoretical works have been focused on the transport and excitation
of magnons contributing to the SSE in the magnetic material \citep{Rezende:2014cr,Cornelissen:2015cz,Kehlberger:2015gg,Cornelissen:2016ji,Kikkawa:2016ck,Barker:2016hy,Basso:2016ii,Adachi:2013jy,Hoffman:2013bl},
whose importance can be recognized in the temperature dependence,
magnetic-field-induced suppression, and thickness dependence of SSEs
\citep{Jaworski:2011ej,Uchida:2012ew,Guo:2016go,Kikkawa:2015bn,Jin:2015ik,Uchida:2014jq}.
Most of the SSE experiments have been performed by using a junction
comprising a ferrimagnetic insulator Y$_{3}$Fe$_{5}$O$_{12}$ (YIG)
and a paramagnetic metal Pt since YIG/Pt enables pure driving and
efficient electric detection of spin-current effects; a YIG/Pt junction
is now recognized as a model system for the SSE studies.

\begin{figure}
\includegraphics{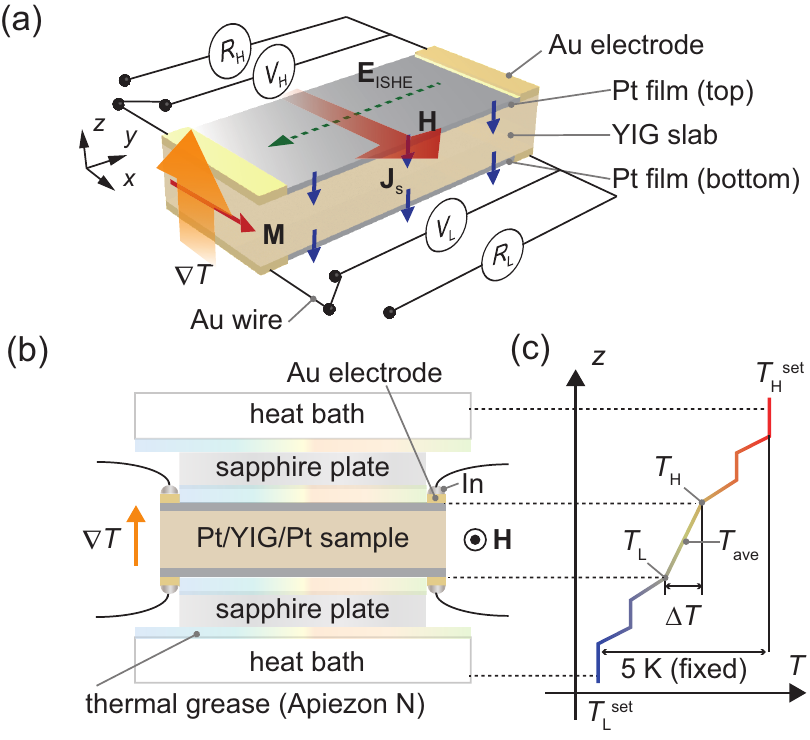} \caption{(a) A schematic illustration of the Pt/YIG/Pt sample. $\nabla T$,
${\bf H}$, ${\bf M}$, ${\bf E}_{{\rm ISHE}}$, and ${\bf J}_{{\rm s}}$
denote the temperature gradient, magnetic field (with the magnitude
$H$), magnetization vector, electric field induced by the ISHE, and
spatial direction of the thermally generated spin current, respectively.
The electric voltage $V_{{\rm H}}$ ($V_{{\rm L}}$) and resistance
$R_{{\rm H}}$ ($R_{{\rm L}}$) between the ends of top (bottom) Pt
layers were measured using a multimeter. (b) Experimental configuration
for applying $\nabla T$. The thickness and width of the sapphire
plates are 0.33 and 2.0 mm, respectively. (c) A schematic plot of
temperature profile along the $z$ direction. }
\label{fig:1} 
\end{figure}

Figure \ref{fig:1}(a) shows a schematic illustration of the SSE in
an YIG/Pt-based system in a longitudinal configuration, which is a
typical configuration used for measuring the SSE. In the longitudinal
SSE (LSSE) configuration, when a temperature gradient $\nabla T$
is applied along the $z$ direction, it generates a spin current across
the YIG/Pt interface \citep{Uchida:2016jo,Uchida:2010jb,Qu:2013io,Xiao:2010iy}.
This thermally-induced spin current is converted into an electric
field ($\mathbf{E}_{{\rm ISHE}}$) by the inverse spin Hall effect
(ISHE) in Pt according to the relation \citep{Saitoh:2006kk,Azevedo:2005kw}
\begin{equation}
{\bf E}_{{\rm ISHE}}\propto{\bf J}_{{\rm s}}\times{\bm{\sigma}},\label{equ:SSE1}
\end{equation}
where ${\bf J}_{{\rm s}}$ is the spatial direction of the spin current
and ${\bm{\sigma}}$ is the spin-polarization vector of $\mathbf{J}_{{\rm s}}$,
which is parallel to the magnetization ${\bf M}$ of YIG {[}see Fig.
\ref{fig:1}(a){]}. When ${\bf M}$ is along the $x$ direction, the
LSSE is detected as a voltage, $V_{{\rm LSSE}}=\int E_{{\rm ISHE}}dy$,
between the ends of the Pt layer along the $y$ direction.

In the LSSE research, temperature dependence of the voltage generation
has been essential for investigating its mechanisms, such as spectral
non-uniformity of magnon contributions \citep{Jin:2015ik,Kikkawa:2015bn,Uchida:2014jq}
and phonon-mediated effects \citep{Adachi:2010ba,Uchida:2012ew,Jaworski:2011ej}.
The recent studies demonstrated that the LSSE voltage in a single-crystalline
YIG slab exhibits a peak at a low temperature, and the peak temperature
is different from that of the thermal conductivity of YIG \citep{Kikkawa:2015bn,Guo:2016go,Jin:2015ik,Boona:2014fh}.
The difference in the peak temperatures was the basis of the recently-proposed
scenario that the LSSE is due purely to the thermal magnon excitation,
rather than the phonon-mediated magnon excitation \citep{Cornelissen:2016ji,Rezende:2014cr,Basso:2016ii}.

In this paper, we report temperature dependence of LSSE in an YIG/Pt-based
system free from thermal artifacts and its strong correlation with
the thermal conductivity of YIG. The LSSE signal and thermal conductivity
were simultaneously measured without changing the experimental configuration.
The intrinsic temperature dependence of the LSSE shows significant
difference from the prior results, while the thermal conductivity
shows good agreement with the previous studies; the peak temperatures
are found to be exceedingly close to each other. These data will be
useful for comparing experiments and theories quantitatively and for
developing comprehensive theoretical models for the SSE.

\label{sec:procedure} The quantitative measurements of the thermoelectric
properties are realized by directly monitoring the temperature difference
between the top and bottom surfaces of a single-crystalline YIG slab.
To do this, we extended a method proposed in Ref. \citealp{Uchida:2014jq},
which is based on the resistance measurements of Pt layers covering
the top and bottom surfaces of the slab {[}Fig. \ref{fig:1}(a){]}.
The lengths of the YIG slab along the $x$, $y$, and $z$ directions
($w$, $l$, and $t$) are 1.9 mm, 6.0 mm, and 1.0 mm, respectively.
After polishing the $x$-$y$ surfaces {[}(111) plane{]} of the YIG
slab, the 10-nm-thick Pt films were sputtered on the whole of the
surfaces. The Pt films are electrically insulated from each other
\citep{Kajiwara:2010ff}. Au electrodes were formed on the edges of
the Pt layers, of which the gap length $l'$ is 5.0 mm. The thermoelectric
voltage and resistance inside the gap were measured by a multimeter.
The sample is put between heat baths with two sapphire plates with
a length of $l'$ for electrical insulation. For thermal connection
between them, thermal grease is used {[}Fig. \ref{fig:1}(b){]}. During
the LSSE measurements, we set $T_{{\rm H}}^{{\rm set}}=T_{{\rm L}}^{{\rm set}}+5~\textrm{K}$
with $T_{{\rm H(L)}}^{{\rm set}}$ being the temperature of the top
(bottom) heat bath (hereafter, we use the subscripts H and L to represent
the corresponding quantities of the top and bottom Pt films, respectively).
In this condition, we monitored the temperature difference between
the top and bottom of the sample $\Delta T=T_{{\rm H}}-T_{{\rm L}}$
by using the Pt films not only as spin-current detectors but also
as temperature sensors {[}Fig. \ref{fig:1}(c){]} \citep{Uchida:2014jq}\footnote{See Supplemental Material attached for details of the temperature
estimation based on the resistance measurements} . A magnetic field with the magnitude $H$ is applied in the $x$
direction.

Importantly, the above method allows us to estimate the thermal conductivity
$\kappa_{{\rm YIG}}$ of the YIG slab at the same time as the LSSE
measurements. This is realized simply by recording the heater power
$P_{{\rm Heater}}$ in addition to $\Delta T$; $\kappa_{{\rm YIG}}$
can be calculated as 
\begin{equation}
\kappa_{{\rm YIG}}=\frac{t}{wl}\frac{P_{{\rm Heater}}}{f\left(l'\right)\Delta T}.\label{equ:thermalconductivity}
\end{equation}
in a similar manner to the steady heat-flow method, where $f\left(l'\right)=0.88$
denotes a form factor for adjusting the measured $\Delta T$ to the
temperature difference averaged over the sample length \footnote{See Supplemental Material attached for details of the $\kappa_{{\rm YIG}}$
measurements}. Note that the influence of the thermal resistances of the Pt layers
\citep{doi:10.1063/1.1378340} and YIG/Pt interfaces \citep{Schreier:2013fz}
are negligibly small for the $\kappa_{{\rm YIG}}$ estimation. The
simultaneous measurements enable quantitative comparison of the LSSE
and $\kappa_{{\rm YIG}}$.

\begin{figure}
\includegraphics{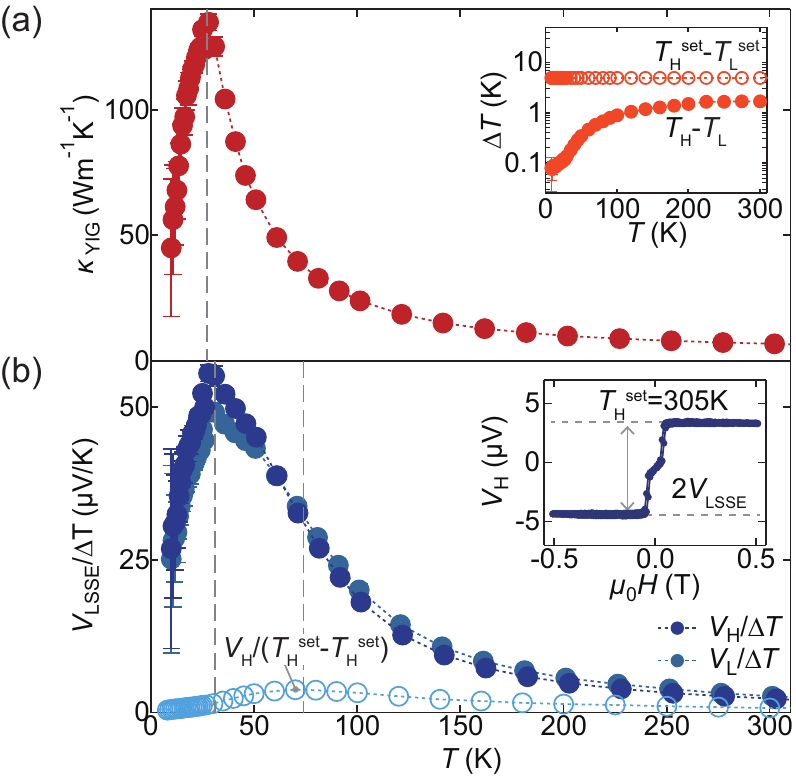}

\caption{Temperature ($T$) dependence of the thermal conductivity $\kappa_{{\rm YIG}}$
estimated from $\Delta T$ and the heater power $P_{{\rm Heater}}$
(a) , and $V_{{\rm LSSE}}/\Delta T$ (b). The inset to (a) shows $\Delta T$
estimated from $R_{{\rm H}}$ and $R_{{\rm L}}$ under the temperature
gradient. The inset to (b) shows $H$ dependence of $V_{{\rm H}}$
at $T_{{\rm H}({\rm L})}^{{\rm set}}=305$ (300) K. The peak temperatures
are determined by parabolic fitting of five points around the maximums.}
\textbf{\label{fig:2}} 
\end{figure}

The inset to Fig. \ref{fig:2}(a) shows the estimated values of $\Delta T$
based on the resistance measurements. We found that the $\Delta T$
value is smaller than the temperature difference applied to the heat
baths ($T_{{\rm H}}^{{\rm set}}-T_{{\rm L}}^{{\rm set}}=5~\textrm{K}$
in this study) at each temperature and strongly decreases with decreasing
the temperature. This behavior can be explained by dominant consumption
of the applied temperature difference by the thermal grease layers;
typical thermal resistance of 10-$\mu$m-thick thermal grease is comparable
to that of the YIG slab at 300 K and much greater than that at low
temperatures as $\kappa_{{\rm YIG}}$ increases at low temperatures
\citep{Boona:2014fh}\footnote{For thermal grease, thermal conductivity of 0.2 ${\rm Wm^{-1}K^{-1}}$
is assumed.}. Consequently, the actual temperature difference ($\Delta T$), and
resultant $V_{{\rm LSSE}}$ measured with fixing $T_{{\rm H}}^{{\rm set}}-T_{{\rm L}}^{{\rm set}}$,
strongly decrease. This result indicates that the conventional method,
which monitors only $T_{{\rm H}}^{{\rm set}}-T_{{\rm L}}^{{\rm set}}$,
cannot reach the intrinsic temperature dependence of the LSSE.

Figure \ref{fig:2}(a) shows $\kappa_{{\rm YIG}}$ as a function of
$T_{{\rm ave}}=\left(T_{{\rm H}}+T_{{\rm L}}\right)/2$, estimated
from Eq. (\ref{equ:thermalconductivity}). The temperature dependence
and magnitude of $\kappa_{{\rm YIG}}$ are well consistent with the
previous studies \citep{Boona:2014fh}, supporting the validity of
our estimation. The $\kappa_{{\rm YIG}}$ value exhibits a peak at
around 27 K and reaches $1.3\times10^{2}$ ${\rm Wm^{-1}K^{-1}}$
at the peak temperature; this temperature dependence can be related
to phonon transport, i.e., the competition between the increase of
the phonon life time due to the suppression of Umklapp scattering
and the decrease of the phonon number with decreasing the temperature
\citep{Boona:2014fh}.

The inset to Fig. \ref{fig:2}(b) shows the $H$ dependence of $V_{{\rm H}}$
in the Pt/YIG/Pt sample at $T_{{\rm H(L)}}^{{\rm set}}=305~\textrm{K}~(300~\textrm{K})$.
The clear LSSE voltage was observed; the voltage shows a sign reversal
in response to the reversal of the magnetization direction of the
YIG slab \citep{Uchida:2016jo,Uchida:2010jb}. We extracted the LSSE
voltage $V_{{\rm LSSE}}$ from the averaged values of $V_{{\rm L}}$
($V_{{\rm H}}$) in the region of $0.2\ {\rm T}<\left|\mu_{0}H\right|<0.5\ {\rm T}$,
where $\mu_{0}$ denotes the vacuum permittivity (note that the field-induced
suppression of the LSSE is negligibly small in this $H$ range \citep{Kikkawa:2015bn,Guo:2016go,Jin:2015ik}).

Figure \ref{fig:2}(b) shows the LSSE voltage normalized by the estimated
temperature difference $V_{{\rm LSSE}}$/$\Delta T$ in the top (bottom)
Pt layers of the Pt/YIG/Pt sample as a function of $T_{{\rm H}}$
($T_{{\rm L}}$). With decreasing the temperature, the $V_{{\rm LSSE}}/\Delta T$
value first increases and, after taking a peak around 31 K, monotonically
decreases toward zero. Although this behavior is qualitatively consistent
with the previous reports \citep{Kikkawa:2015bn,Guo:2016go,Jin:2015ik,Boona:2014fh},
the observed peak structure is very steep and the peak temperature
significantly differs from the previous results, where the peak temperature
was reported to be $\sim70$ K. We found that, if $V_{{\rm LSSE}}$
is normalized by the temperature difference between the heat baths
$T_{{\rm H}}^{{\rm set}}-T_{{\rm L}}^{{\rm set}}$, our data also
exhibits a peak $\sim$74 K as previously reported \citep{Kikkawa:2015bn,Guo:2016go,Jin:2015ik,Boona:2014fh}.
The overestimated peak temperature is attributed to the misestimation
of $\Delta T$ shown in the inset to Fig. \ref{fig:2}(a), showing
that the previous results may not represent intrinsic temperature
dependence of the LSSE because of the thermal artifacts, and the direct
$\Delta T$ monitoring is necessary for the quantitative LSSE measurements.
We also note that the magnitude of the intrinsic LSSE thermopower
is in fact much greater than that estimated from the previous experiments.
The difference in the signal magnitude is more visible at low temperatures;
the LSSE voltage in our Pt/YIG/Pt sample at the peak temperature reaches
to $V_{{\rm LSSE}}/\Delta T\sim55\ \mu\textrm{V/K}$.

\begin{figure*}
\includegraphics{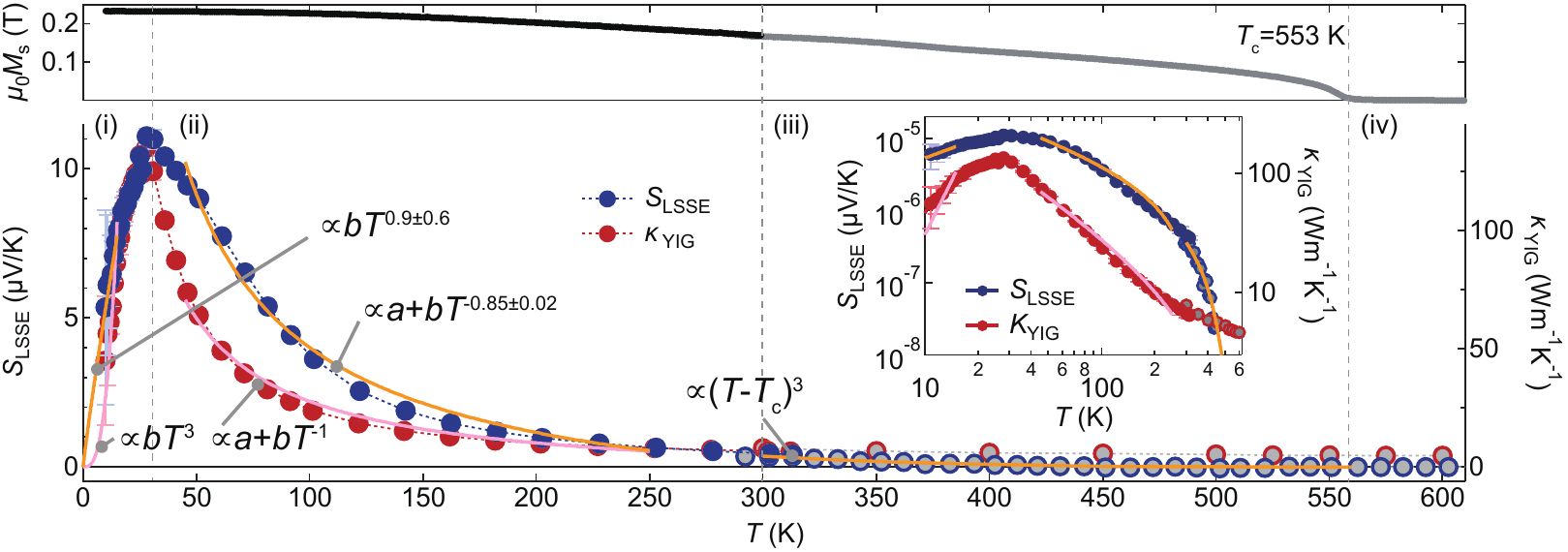} \caption{The saturation magnetization $M_{{\rm s}}$ of the YIG slab as a function
of $T$, the LSSE thermoelectric coefficient $S_{{\rm LSSE}}$ of
the top Pt layer as a function of $T_{{\rm H}}$, and the thermal
conductivity $\kappa_{{\rm YIG}}$ as a function of the averaged temperature
$T_{{\rm ave}}$. $M_{{\rm s}}$ was measured using a vibrating sample
magnetometer. The data at temperatures higher than 300 K is drawn
from Ref. \citealp{Uchida:2014jq}, which shares the same YIG slab
and Pt thickness. The inset shows a Log-Log plot. The solid lines
represent fitting curves discussed in the main text. }

\label{fig:3} 
\end{figure*}

In Fig. 3, we show the complete temperature dependence of the LSSE
thermopower $S_{{\rm LSSE}}=tV_{{\rm LSSE}}/\left(l'\Delta T\right)$,
thermal conductivity $\kappa_{{\rm YIG}}$, and saturation magnetization
$M_{{\rm s}}$ of the YIG, which includes the data in the high temperature
range reported in Ref. \citealp{Uchida:2014jq}. In the following,
we discuss the origin of the temperature dependence of the LSSE. The
temperature dependence of the LSSE originates from the spin mixing
conductance \citep{Tserkovnyak:2005fr,Zhang:2012hh} at Pt/YIG interfaces,
spin diffusion length, spin Hall angle \citep{maekawa2012spin}, and
resistance of Pt, and dynamics of thermally-excited magnons in YIG.
The spin mixing conductance, which is proportional to the LSSE voltage,
may depend on $T$ and $M_{{\rm s}}$. The predicted relation, $\propto M_{{\rm s}}^{2}$
\citep{Ohnuma:2014cc}, cannot explain the LSSE enhancement at low
temperatures because the maximum possible enhancement is calculated
as a factor of 1.9. Similarly, the spin diffusion length or spin Hall
angle of Pt cannot explain the enhancement of the LSSE voltage at
low temperatures as they are almost temperature-independent below
300 K \citep{Vila:2007jn,Sagasta:2016iv}. The temperature dependence
of the resistance of the Pt is also irrelevant because it decreases
with decreasing the temperature. Therefore, the LSSE enhancement at
the peak likely comes from the properties of thermally-excited magnons.
Here, two scenarios have been proposed for magnon excitation in the
SSE: one is the pure thermal magnon excitation \citep{Cornelissen:2016ji,Rezende:2014cr,Guo:2016go,Basso:2016ii,Jin:2015ik}
and the other is the phonon-mediated magnon excitation \citep{Adachi:2010ba,Jaworski:2010dy,Uchida:2011bc}.

Importantly, the peak temperature of the observed LSSE signal in our
Pt/YIG/Pt sample is almost the same as that of $\kappa_{{\rm YIG}}$,
indicating the strong correlation between $V_{{\rm LSSE}}$ and $\kappa_{{\rm YIG}}$
at low temperatures. This behavior is consistent with the scenario
of the phonon-mediated SSE, where SSE voltage is expected to be proportional
to the phonon life time in YIG (note again that the peak in the $T$-$\kappa_{{\rm YIG}}$
curve reflects the phonon transport) \citep{Adachi:2010ba}. Although
the recent studies proposed the scenario that the LSSE is due purely
to the magnon-driven contribution, it is based on the difference in
the peak temperatures between the LSSE and thermal conductivity, which
now turned out to be relevant to the thermal artifacts {[}see Fig.
2(b){]}. The magnon- and phonon-driven contributions cannot be separated
completely by the present experiments. However, at least, the presence
of the phonon-mediated process cannot be excluded because of the similar
peak temperatures between $S_{{\rm LSSE}}$ and $\kappa_{{\rm YIG}}$.

According to the previous studies on the transverse SSE \citep{Jaworski:2011ej},
the observed temperature dependence of the LSSE can be separated into
the following four regions: (i) the low temperature region from 10
to 30 K, (ii) from 30 K to room temperature, (iii) the high temperature
region from room temperature to the Curie temperature $T_{{\rm c}}$
of YIG, and (iv) above $T_{{\rm c}}$. In the region (i), both $S_{{\rm LSSE}}$
and $\kappa_{{\rm YIG}}$ increase with the temperature $T$ and then
reach their maximums. This tendency can be expressed by $bT^{\gamma}$
with $\gamma=0.9\pm0.6$ for $S_{{\rm LSSE}}$ up to 15 K, where $b$
and $\gamma$ are fitting parameters. $\kappa_{{\rm YIG}}$ in this
temperature range is reproduced by setting $\gamma=3$ \citep{Boona:2014fh}
. The difference in the $\gamma$ exponents between $S_{{\rm LSSE}}$
and $\kappa_{{\rm YIG}}$ can be due to difference in the $T$ dependence
of the heat capacitance and group velocity between magnons and phonons
\citep{Jaworski:2010dy,Boona:2014fh,Jin:2015ik}. In the region (ii),
$S_{{\rm LSSE}}$ and $\kappa_{{\rm YIG}}$ start to decrease with
increasing $T$. While $\kappa_{{\rm YIG}}$ shows $a+bT^{-1}$ dependence
originating from Umklapp scattering of phonons, $S_{{\rm LSSE}}$
shows the $a+bT^{-0.85\pm0.02}$ dependence. This difference suggests
the coexistence of the magnon- and phonon-induced processes in the
LSSE, reflecting different temperature dependence of the magnon and
phonon life times \citep{Boona:2014fh,Rezende:2014if}. In the region
(iii), $S_{{\rm LSSE}}$ shows strong correlation to $M_{{\rm s}}$
rather than $\kappa_{{\rm YIG}}$, differently from (i). Here, $S_{{\rm LSSE}}$
and $M_{{\rm s}}$ are described by $\left(T_{{\rm c}}-T\right)^{\gamma}$
with $\gamma=3$ and 0.5, respectively \citep{Uchida:2014jq}, while
$\kappa_{{\rm YIG}}$ gradually decreases. Finally, in the region
(iv), $S_{{\rm LSSE}}$ and $M_{{\rm s}}$ vanish. The microscopic
and quantitative explanation of the above behavior remains to be achieved.

\begin{figure}
\includegraphics{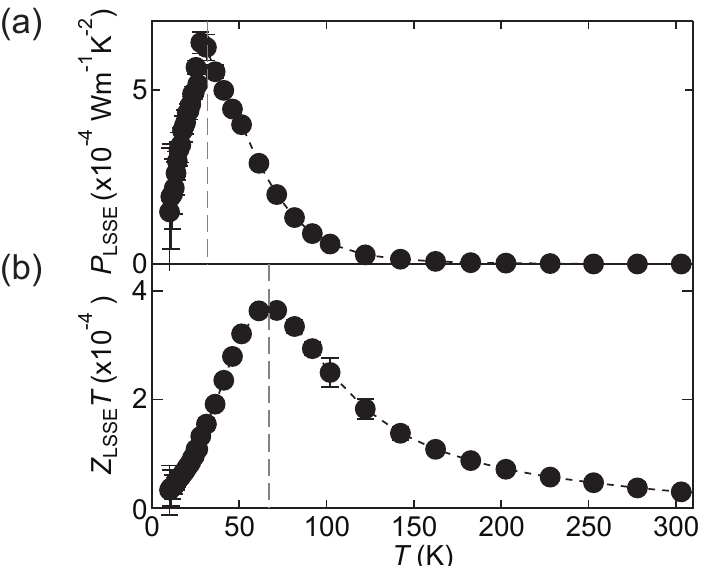} \caption{$T$ dependence of the power factor $P_{{\rm LSSE}}$ (a) and the
figure of merit $Z_{{\rm LSSE}}T$ of our LSSE device (b).}
\label{fig:4} 
\end{figure}

Since the experimental method demonstrated here enables simultaneous
measurements of the LSSE thermopower $S_{{\rm LSSE}}$, thermal conductivity
of YIG $\kappa_{{\rm YIG}}$, and electrical conductivity of Pt $\sigma_{{\rm Pt}}$,
we can also obtain the quantitative temperature dependence of the
thermoelectric performance of our Pt/YIG/Pt sample. Figure 4(a) shows
the power factor $P_{{\rm LSSE}}=S_{{\rm LSSE}}^{2}\sigma_{{\rm Pt}}$
as a function of $T$. Owing to the strong enhancement of $S_{{\rm LSSE}}$
at low temperatures, the power factor exhibits a sharp peak at 32
K; the $P_{{\rm LSSE}}$ value at the peak temperature is 6$\times10^{-4}$
${\rm Wm^{-1}K^{-2}}$, $\sim1000$ times greater than that at room
temperature. In contrast, the figure of merit $Z_{{\rm LSSE}}T=(S_{{\rm LSSE}}^{2}\sigma_{{\rm Pt}}/\kappa_{{\rm YIG}})T$
\citep{Uchida:2016jo} exhibits a maximum at 67 K due to the competition
between $S_{{\rm LSSE}}$ and $\kappa_{{\rm YIG}}$ {[}Fig. 4(b){]}.
The low-temperature enhancement of the power factor and figure of
merit is in sharp contrast to the typical behavior of the conventional
Seebeck devices \citep{Zhang201592}, the thermoelectric performance
of which decreases with decreasing the temperature, indicating potential
thermoelectric applications of LSSE at low temperatures.

In conclusion, we systematically investigated the longitudinal spin
Seebeck effect (LSSE) in an Y$_{3}$Fe$_{5}$O$_{12}$ (YIG) slab
sandwiched by two Pt films in the low temperature range from 10 K
to room temperature. The direct temperature monitoring based on the
resistance measurements of the Pt layers successfully reveals the
intrinsic LSSE behavior, unreachable by the conventional method. We
found that the magnitude of the LSSE in the Pt/YIG/Pt sample rapidly
increases with decreasing temperature and takes a maximum at a temperature
very close to the peak temperature of the thermal conductivity of
YIG. The strong correlation between the LSSE and thermal conductivity
shed light again on the importance of the phonon-mediated processes
in the SSE. Although more detailed experimental and theoretical investigations
are required, we anticipate that the finding of the intrinsic temperature
dependence of the LSSE will be helpful for obtaining the full understanding
of its mechanism.
\begin{acknowledgments}
The authors thank J. Shiomi, A. Miura, T. Oyake, H. Adachi, T. Kikkawa,
T. Ota, R. Ramos, and G. E. W. Bauer for valuable discussions. This
work was supported by PRESTO ``Phase Interfaces for Highly Efficient
Energy Utilization'' and ERATO ``Spin Quantum Rectification'' from
JST, Japan, Grant-in-Aid for Scientific Research (A) (JP15H02012),
Grant-in-Aid for Scientific Research on Innovative Area ``Nano Spin
Conversion Science'' (JP26103005) from JSPS KAKENHI, Japan, NEC Corporation,
the Noguchi Institute, and E-IMR, Tohoku University. S.D. is supported
by JSPS through a research fellowship for young scientists (No. 16J02422).
\end{acknowledgments}

%%\bibliography{lowtempbib}

\newpage{}

\section*{Supplemental Materials}

\def\thesection{S. \arabic{section}} 
\def\thefigure{S\arabic{figure}} 
\def\theequation{S\arabic{equation}} 
\setcounter{figure}{0}
\setcounter{equation}{0}

\section{Temperature estimation}

To estimate the temperature difference of the sample, we monitored
the temperatures $T_{{\rm H}}$ and $T_{{\rm L}}$ of the Pt layers
on the top and bottom surfaces of the YIG slab based on the resistance
measurements as in Ref. \cite{Uchida:2014jq}. In this experiment,
the series resistance of the Pt layer, 0.5-mm-long Au electrodes,
and Au wires were measured using a four probe method. The Au wires
with a diameter of 50 $\mu$m were rigidly connected to the Au electrodes
on the Pt/YIG/Pt sample via indium soldering. Since the electrodes
and wires have negligibly small resistance, the measured values and
temperatures can be attributed to those in the region $-l'/2<y<l'/2$
of the Pt layer, where the gap length $l'$ is 5 mm in this study
and $y=0$ is set to the center of the YIG slab. This does not affect
to the LSSE estimation, because the output voltage only appears in
the Pt layers in the area without the Au electrodes, while it should
be considered in the thermal-conductivity estimation (as discussed
in the next section).

The temperature of the Pt layers are determined by comparing the resistances
of the Pt films $R_{{\rm H}}$ and $R_{{\rm L}}$ under the temperature
difference with the isothermal $R_{{\rm H}}$ and $R_{{\rm L}}$ at
various temperatures. The LSSE measurements started from 8 K to 300
K. The resistance measurements were performed three times at each
temperature: (1) at the isothermal condition before the LSSE measurements,
(2) at the steady-state condition with the heater output on ($T_{{\rm H}}^{{\rm set}}-T_{{\rm L}}^{{\rm set}}=5.0~\textrm{K}$),
and (3) again at the isothermal condition after the heater is turned
off. The $R_{{\rm H(L)}}$ values recorded in the process (2) were
transformed into the $T_{{\rm H(L)}}$ by comparing them with the
isothermal $R_{{\rm H(L)}}$ values. Immediately after the process
(2), the magnetic field $H$ dependence of the voltages $V_{{\rm H}}$
and $V_{{\rm L}}$, the LSSE voltages, was measured. After the LSSE
measurement, the process (3) was performed to check the reproducibility
of the resistance. The current amplitude for sensing $R_{{\rm H}}$
and $R_{{\rm L}}$ is 1 mA, which induces sufficiently small heat
{[}0.1 \% of typical applied heater output ($P_{{\rm Heater}}$){]},
and does not affect the original sample temperatures.

\begin{figure}
\includegraphics{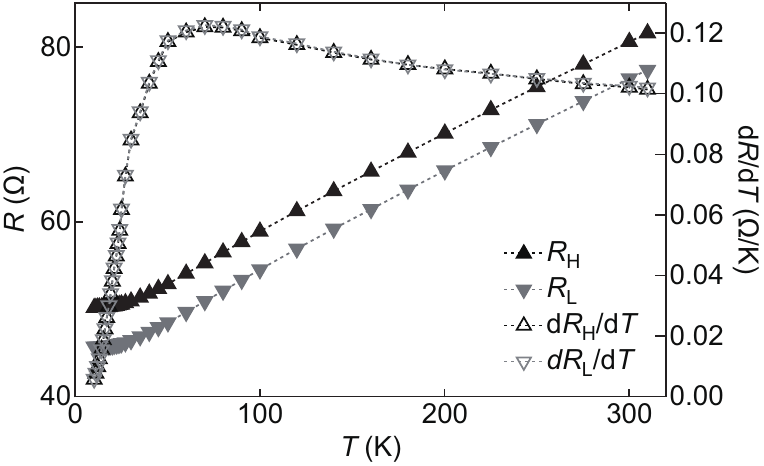}

\caption{Temperature $T$ dependence of $R_{{\rm H}}$, $R_{{\rm L}}$, and
their differentials.}

\label{fig:s0}
\end{figure}

Figure \ref{fig:s0} shows the temperature ($T$) dependence of $R_{{\rm H}}$
and $R_{{\rm L}}$ of the Pt films measured under the isothermal conditions
and their differentials. As the $R_{{\rm H}({\rm L})}-T$ curve is
monotonic above 10 K, we confined the measurements in the range from
10 to 300 K, where the sufficient temperature sensitivity is ensured
with our measurement accuracy. The difference in the resistance between
before and after the LSSE measurements was $<1\times10^{-3}\ \Omega$
below 100 K, corresponding to the error of $<0.01$ K except 10\textendash 13
K. The standard deviation of the typical resistance measurements is
about $1\times10^{-4}\ \Omega$. The difference values and the standard
deviation of the resistance values were treated as the error of the
temperature estimation, which contributes to the error bars in Fig.
2\textendash 4. 

\section{Thermal conductivity estimation}

\begin{figure*}[!tp]
\includegraphics{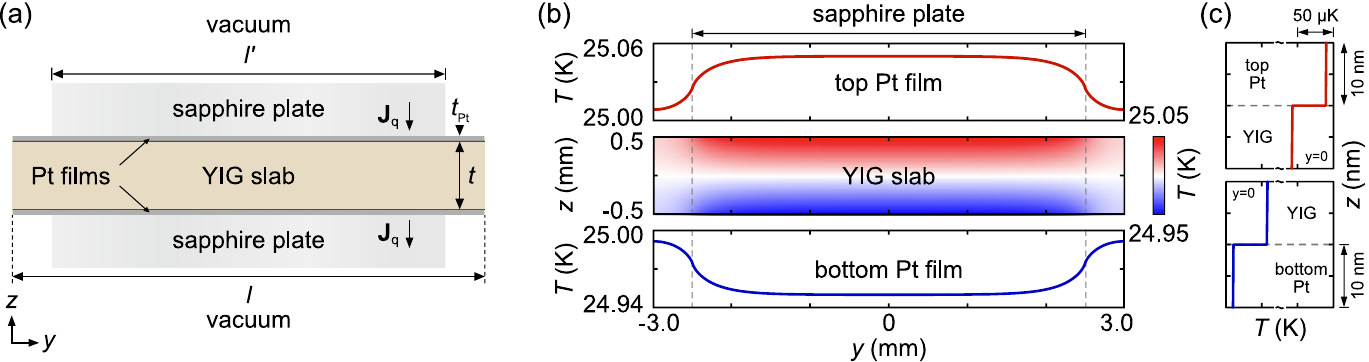}

\caption{(a) Model system for temperature distribution calculation. YIG slab
with a length ($l$) of 6.0 mm and a thickness ($t$) of 1.0 mm is
sandwiched by two 10-nm-thick Pt layers and two sapphire plates with
the gap length $l'=5.0$ mm. The heat current $J_{q}$ flows from
the top sapphire plate. (b) Calculated temperature distribution in
the system. Because of the size difference of the sample and the input
heat flow, the heat distribution becomes non uniform. The averaged
temperature difference $\Delta\tilde{T}$ over $l$ differs from the
measured averaged temperature $\Delta T$, which is averaged over
$-l'/2<y<l'/2$. (c) Temperature profile along the thickness direction
(the $z$ axis) at the center of the $x-y$ plane. \label{fig:S1}}
\end{figure*}

The thermal conductivity $\kappa$ of the system is calculated based
on the thermal diffusion equation in the steady state condition, $\nabla^{2}T=0$.
We consider a model system shown in Fig. \ref{fig:S1}(a). By considering
the temperature distribution symmetric with respect to the $z$ axis,
we obtain a thermal diffusion equation for the averaged temperature
$\tilde{T}\left(z\right)$ over the $y$ direction in the Pt/YIG/Pt
system,
\begin{equation}
\partial_{z}^{2}\tilde{T}\left(z\right)=0.\label{eq:th2}
\end{equation}
In this condition, $\kappa$ can be obtained as 
\begin{equation}
\kappa=\frac{t+2t_{{\rm Pt}}}{wl}\frac{P}{\Delta\tilde{T}}\label{eq:kappa}
\end{equation}
with $\Delta\tilde{T}=\tilde{T}\left(t/2+t_{{\rm Pt}}\right)-\tilde{T}\left(-t/2-t_{{\rm Pt}}\right)$
under the boundary conditions: 
\begin{equation}
\kappa\partial_{z}\tilde{T}\left(\pm\left[t/2+t_{{\rm Pt}}\right]\right)=\mp\frac{J_{q}}{lw},\label{eq:}
\end{equation}
where $t$ ($t_{{\rm Pt}}$) denotes the thickness of the YIG slab
(the Pt layer) $w$ the width of the sample, and $J_{q}=P_{{\rm Heater}}$
the applied heat current. Here, $z=0$ is the center of the sample.
This equation is valid for nonuniform heat flows when the entire heat
goes through the sample and sapphire plates. 
Since our estimated $\Delta T$, which is the averaged temperature
difference over $-l'/2<y<l'/2$, slightly differs from $\Delta\tilde{T}$,
we need a form factor, $f\left(l'\right)=\Delta\tilde{T}/\Delta T$,
calculated by a numerical calculation of the heat distribution using
the COMSOL software. The difference is due to the existence of the
conductive Au electrodes and the nonuniform heat-flow due to the size
difference between the input heat current and the sample dimension
{[} See Fig. \ref{fig:S1}(a){]}. Figure \ref{fig:S1}(b) shows the
calculated 2D temperature profile in the sample, where we assumed
that the thermal conductivity of YIG and Pt is 1.3$\times10^{2}$
${\rm Wm^{-1}K^{-1}}$ (the maximum value at $\sim27\ {\rm K}$ in
our experiment) and $20\times10$ ${\rm Wm^{-1}K^{-1}}$ \cite{doi:10.1063/1.1378340},
respectively, and the interfacial thermal conductance of the YIG/Pt
is 2.79$\times10^{8}$ ${\rm Wm^{-2}K^{-1}}$ \cite{Schreier:2013fz}.
The calculated form factor becomes $f\left(l'\right)=0.88$ and is
temperature independent. 

In our system, $\kappa$ can be regarded as that of the YIG slab because
of the small values of $t_{{\rm Pt}}$ and the interfacial thermal
resistance of the YIG/Pt interface. Fig. \ref{fig:S1}(c) shows the
temperature distribution along the $z$ axis assuming $T=28$ K, at
which the thermal resistance of the YIG slab is minimized and thus
the effect of the film and interface may be maximized. As expected,
the temperature drop in the Pt layers is about $7\times10^{-4}$ \%
of that in the YIG slab, and the interfacial temperature drop is about
$5\times10^{-2}$ \%, negligibly small contributions. 

\begin{figure}
\includegraphics{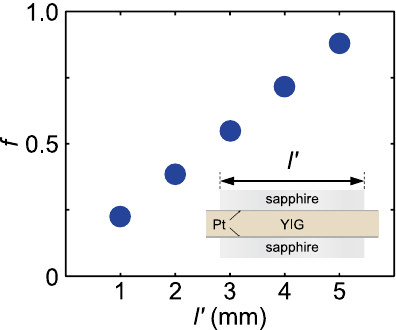}\caption{Form factor $f\left(l'\right)$ for various $l'$ values. Here, the
other dimensions are the same as that used for the calculation of
Fig. \ref{fig:S1}.\label{fig:S2}}
\end{figure}

For experiments with a sample having another dimension, we calculate
the form factor $f\left(l'\right)$ for various $l'$ values, which
is shown in Fig. \ref{fig:S2}.

\end{document}